\documentclass{article}

\usepackage{PRIMEarxiv}

\usepackage[utf8]{inputenc} 
\usepackage[T1]{fontenc}    
\usepackage{hyperref}       
\usepackage{url}            
\usepackage{booktabs}       
\usepackage{amsfonts}       
\usepackage{nicefrac}       
\usepackage{microtype}      
\usepackage{lipsum}
\usepackage{fancyhdr}       
\usepackage{graphicx}       
\usepackage{lineno}
\usepackage{amsmath}
\usepackage{float}
\usepackage{graphicx}
\usepackage{bm}
\usepackage{multirow}
\usepackage{booktabs}
\usepackage{amsmath}
\usepackage{algorithm}
\usepackage{algpseudocode}  
\usepackage{amsfonts}
\makeatletter
\renewcommand{\ALG@beginalgorithmic}{\setlength{\itemsep}{1.4ex}} 

\graphicspath{{media/}}     

\title{Single-pixel imaging via data-driven and deep image prior dual-network
\thanks{\textit{\underline{Citation}}: 
\textbf{Jing-Yi Shi, et al, "Single-pixel imaging via data-driven and deep image prior dual networks," Opt. Express 33, 26690-26702 (2025) DOI:https://doi.org/10.1364/OE.563184.}}
}
\author{
Jing-Yi Shi\textsuperscript{1,2}\quad
Jia-Qi Song\textsuperscript{1} \quad
Peng-Cheng Ji\textsuperscript{3} \quad
Zi-Qing Zhao\textsuperscript{1,2} \quad
Yuan-Jin Yu\textsuperscript{3} \And
Ming-Fei Li\textsuperscript{1,2} \thanks{Corresponding author. \texttt{mf\_li@iphy.ac.cn}} \quad 
Ling-An Wu\textsuperscript{1,2} \quad 
\\
\\
\textsuperscript{1}\,Institute of Physics, China Academy of Sciences, Beijing, 100190 China \\
\textsuperscript{2}\,School of Physical Sciences, University of Chinese Academy of Sciences, Beijing 100049, China \\
\textsuperscript{3}\,School of Automation, Beijing Institute of Technology, Beijing, 100081 China
}

\begin{document}
\maketitle

\begin{abstract}
Single-pixel imaging (SPI), especially when integrated with deep neural networks like Deep Image Prior Networks (DIP-Net) or Data-Driven Networks (DD-Net), has gained considerable attention for its capability to generate high-quality reconstructed images, even in the presence of sub-sampling conditions. However, DIP-Net often requires thousands of iterations to achieve high-quality image reconstruction, and DD-Net performs optimally only when the target closely resembles the features present in its training set. To overcome these limitations, we propose a Dual-Network Iterative Optimization (SPI-DNIO) framework that combines the strengths of both DD-Net and DIP-Net. It has been demonstrated that this approach can recover high-quality images with fewer iteration steps. Furthermore, to address the challenge of SPI inputs having less effective information at low sampling rates, we have designed a residual block enriched with gradient information, which can convey details to deeper layers, thereby enhancing the deep network's learning capabilities. We have applied these techniques to both indoor experiments with active lighting and outdoor long-range experiments with passive lighting. Our experimental results confirm the exceptional reconstruction capabilities and generalization performance of the SPI-DNIO framework.
\end{abstract}


\section{Introduction}
Traditional imaging schemes capture light that is either reflected or scattered by an object to directly form an image, embodying the principle of "what you see is what you get." In contrast, single-pixel imaging (SPI) separates the imaging process into two distinct phases: detection (encoding) and reconstruction (decoding). During the detection (encoding) phase, a series of spatial patterns are used to illuminate the object, followed by the employment of a bucket detector for one-dimensional (1D) signal acquisition. In the reconstruction (decoding) phase, a two-dimensional (2D) image is formed by leveraging the second-order correlations of classical light or through the application of inversion algorithms \cite{1,2,3,4}. Furthermore, SPI utilizes a bucket detector to collect the majority of photons interacting with the objects. This approach endows SPI with exceptional performance in several aspects, such as faster timing responses, reduced dark counts, and enhanced detection sensitivity \cite{5}. These advances have sparked interest in applying SPI across various fields, including THz Imaging \cite{6,7}, 3D imaging \cite{8,9}, and remote sensing \cite{10}, and other applicatoins \cite{11,12} over the past decade. However, in SPI, if the spatial patterns used for 1D measurements are non-orthogonal or affected by noise, a large number of such measurements are required to reconstruct an image with good quality  \cite{13}. Obviously, designing the encoding patterns with orthogonality \cite{4,14,15,16,17} or optimizing the patterns \cite{18,19,20} that ensure each measurement contains as much effective information as possible is an effective solution to overcome this limitation. Additionally, compressive sensing (CS) \cite{21} offers another solution by significantly reducing the number of required patterns and measurements in SPI, capitalizing on the sparsity of natural images in specific measurement bases. Subsequently, scholars have developed optimization algorithms based on CS to achieve high reconstruction quality with fewer measurements \cite{22,23,24}. However, typical CS algorithms utilize the framework of iterative convex optimization for image reconstruction, which requires substantial computational time \cite{25,26}. Meanwhile, traditional CS algorithms often struggle to achieve good image quality for low resolutions when the sampling ratio is below 30\% \cite{14}.

Recently, deep learning (DL) due to its powerful image compression and reconstruction capabilities, is adopted to SPI \cite{27,28,29}. The most direct adoption is utilizing an end-to-end deep-learning approach, which can make the deep neural network (DNN) learn a mapping relationship from a large number of input-output data pairs. This method can recover the object directly from the detected bucket signal \cite{29}. Meanwhile, integrating physical information into DNN training can increase the network's performance and Interpretability \cite{30}. In addition, the untrained DNNs with the deep image prior (DIP) have been demonstrated to exhibit excellent results in solving inverse problems \cite{31,32}. In 2022 years, F. Wang et al proposes a DIP network (DIP-Net) combined with physical information which achieves high-quality reconstruction of ghost imaging (GI) through iterative optimization \cite{33}.

Despite the advancements of single-network frameworks like DIP-Net and data-driven networks (DD-Net) for single-pixel imaging (SPI) reconstruction, both approaches have inherent limitations. For DIP-Net, achieving high-quality reconstructions typically necessitates thousands of iterations, while DD-Net excels in image reconstruction only when the target images share similar characteristics with those in the training set. To address these shortcomings, we propose the Dual-Network Iterative Optimization (SPI-DNIO) framework, which synergistically combines the strengths of both DD-Net and DIP-Net. This approach enables high-performance SPI reconstruction while reducing optimization iterations to a few hundred steps.

Additionally, to mitigate the challenge posed by SPI inputs with limited information at low sampling rates, we introduce a residual block with Rich Gradient Information (RGI). This mechanism facilitates the flow of information to deeper layers, enhancing the network’s learning capacity. To evaluate the generalizability of SPI-DNIO, we conduct simulations across multiple datasets, including LFW \cite{34}, DIV2K \cite{35}, Set14 \cite{36}, and LSDIR \cite{37}. Furthermore, we test SPI-DNIO in both indoor experiments with active lighting and outdoor long-range experiments with passive lighting to assess its robustness. Our results demonstrate SPI-DNIO's outstanding performance in terms of reconstruction quality, generalization, and robustness.

\section{Methods}

In the SPI system, the 1D measurements of the object $T({x_i},y{}_i)$ are acquired through a single-pixel detector that lacks spatial resolution. These measurements ${I_m}$ can be described as
\begin{align}
 \begin{gathered}
{I_m} = \int {{H_m}({x_i},{y_i})T({x_i},{y_i})dxdy}  \propto \sum\limits_{{x_i},{y_i}} {{H_m}({x_i},{y_i})T({x_i},{y_i})}.
  \end{gathered}
 \label{eq:1}
\end{align}
Where, ${H_m}({x_i},{y_i})$ represents the random illumination patterns, with $m$ = 1, 2..., $M$.
In the conventional GI reconstruction algorithm, the object is typically reconstructed by calculating the intensity correlation between ${H_m}$ and ${I_m}$  
\begin{align}
 \begin{gathered}
{O_{GI}} = \left\langle {{H_m}{I_m}} \right\rangle  - \left\langle {{H_m}} \right\rangle \left\langle {{I_m}} \right\rangle,
  \end{gathered}
 \label{eq:2}
\end{align}
where $\left\langle  \right\rangle$ denotes the ensemble average. $\left\langle {{I_m}} \right\rangle  = \frac{1}{M}\sum\limits_{m = 1}^M {{I_m}}$ and $\left\langle {{I_m}} \right\rangle  = \frac{1}{M}\sum\limits_{m = 1}^M {{I_m}} $, respectively. Comparing to GI reconstruction algorithm, the differential GI (DGI)  demonstrated the higher imaging reconstruction quality \cite{38}. The DGI reconstruction algorithm can be described as
\begin{align}
 \begin{gathered}
{O_{DGI}} = \left\langle {{H_m}{I_m}} \right\rangle  - \frac{{\left\langle {{I_m}} \right\rangle }}{{\left\langle {{S_m}} \right\rangle }}\left\langle {{S_m}{H_m}} \right\rangle,
  \end{gathered}
 \label{eq:3}
\end{align}
\begin{figure*}[htpb]
    \centering
    \includegraphics[width=0.9\linewidth]{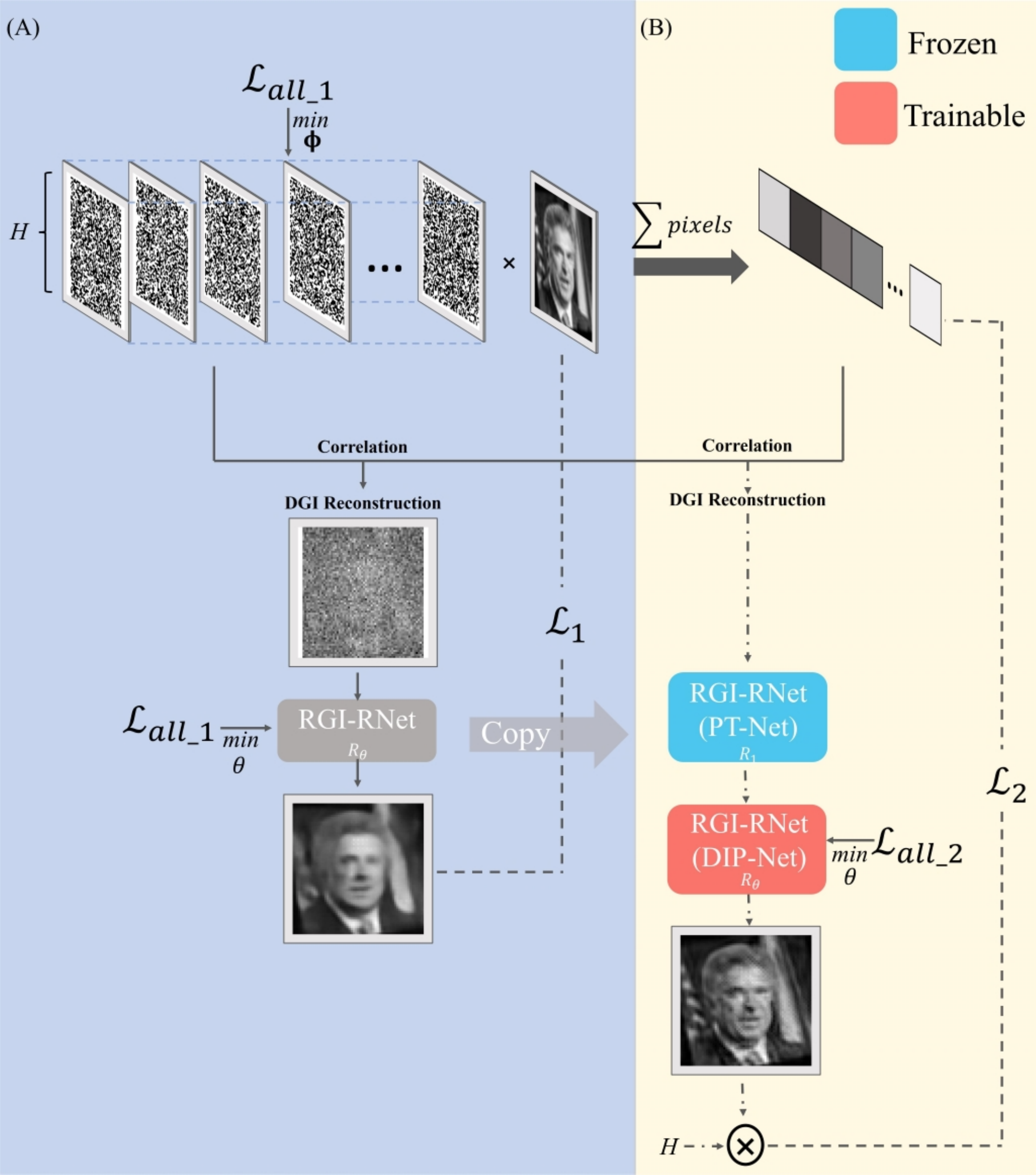}
    \caption{The specific steps of the SPI-DNIO. (A) The process of network training and optimizing patterns. (B) The Iterative optimization for the image reconstruction. Imaging results of the LFW training dataset (Randomly select 10000 images from the LFW dataset) for 64 $\times$ 64 resolution images at 25\% sampling ratio. The face images were taken from LFW\cite{34}}
    \label{fig:1}
\end{figure*}
where ${S_m} = \sum\limits_{{x_i},{y_i}} {{H_m}({x_i},{y_i})}$.
The SPI-DNIO proposed in this work mainly consists of two steps: network training and iterative optimization. In the first step, a three-dimensional tensor $H(h,w,M)$ serving as the optimizable pattern, is randomly initialized, where $(h,w)$ represents the size of the image, and $M$ denotes the actual number of measurements. During the forward propagation, the optimizable pattern computes its gradients, enabling their optimization in the backward propagation.
To incorporate physical constraints into the model training, the optimizable patterns are correlated with the measurements to reconstruct DGI which serves as the network input \cite{30}. 

Specifically, the training process shown in Fig. \ref{fig:1}(A) can be formulated as follows 
\begin{align}
 \begin{gathered}
\{ {{\rm H}^*},{\theta ^*}\}  = \arg {\min _{{\rm H},\theta }}{\mathcal{L}_{all\_1}} = {\mathcal{L}_1}({R_\theta }({O_{DGI}}),T) + F(\boldsymbol{\Phi} ),
  \end{gathered}
 \label{eq:4}
\end{align}
where ${R_\theta }$ refers to the residual network with rich gradient information (RGI-RNet) and $\theta $ represents the parameters of the RGI-RNet, whose corresponding design principles and structures are illustrated in Fig. \ref{fig:3}. ${\mathcal{L}_1}(y,\hat y) = {\left| {y - \hat y} \right|^2}$ represents the reconstruction loss in the first step, where $y$ is the predicted value and $\hat y$ is the ground truth value. $F({\rm \boldsymbol{\Phi}})$ is the physical constraint regularizer for regularizing patterns $\boldsymbol{\Phi} (h \times w,{\rm M}) = \rm {reshape}(\rm H)$ and we set it as follows
\begin{align}
 \begin{gathered}
F(\boldsymbol{\Phi} ) = {\lambda _1}{\sum\limits_{ij} {(1 - {\boldsymbol{\Phi} _{ij}})} ^2}{(1 + {\boldsymbol{\Phi} _{ij}})^2} + {\lambda _2}{\left| {{\rm \boldsymbol{I_M}} - \frac{1}{M}{\boldsymbol{\Phi} ^T}\boldsymbol{\Phi} } \right|^2},
  \end{gathered}
 \label{eq:5}
\end{align}
where ${\lambda _1}$ and ${\lambda _2}$ are the regularization parameters that control the trade-off between the loss function, with their corresponding values set to 0.01 and 1 respectively. ${\lambda _1}{\sum\limits_{ij} {(1 - {\boldsymbol{\Phi} _{ij}})} ^2}{(1 + {\boldsymbol{\Phi} _{ij}})^2}$ promotes ${\boldsymbol{\Phi} _{ij}}$ to assume values of $1$ or $-1$ during its minimization process \cite{39}. ${\left| {{\rm \boldsymbol{I_M}} - \frac{1}{M}{\boldsymbol{\Phi} ^T}\boldsymbol{\Phi} } \right|^2}$ is utilized to improve orthogonalization of $\boldsymbol{\Phi}$. $\rm \boldsymbol{I_M}$ represents the identity matrix and ${\boldsymbol{\Phi} ^T}$ denotes the transpose of $\boldsymbol{\Phi}$. During the end-to-end training process, the network establishes a mapping relationship between the input and output data. Simultaneously, the random patterns are optimized to meet physical constraints. In the second step, both the network parameters and the patterns trained in the first step are frozen, and the second network whose parameters are randomly initialized, is designed with the same structure as the network from the first step.
During the optimizing iteration, the second network is optimized using the loss ${\mathcal{L}_{all\_2}}$ to further enhance its image reconstruction capability. The details of the iterative optimization function are as follows
\begin{align}
 \begin{gathered}
\{ {\theta ^*}\}  = \arg {\min _\theta }{\mathcal{L}_{all\_2}} = \frac{1}{M}\sum\limits_{m = 1}^M {{{\left| {H_m^*{R_\theta }(\tilde O) - {I_m}} \right|}^2}}  + {\lambda _3}\operatorname{TV} (\tilde O),
  \end{gathered}
 \label{eq:6}
\end{align}
where $\tilde O = {R_1}({O_{DGI}})$. ${R_1}()$ is the pre-trained network (PT-Net) in the first step, ${\rm H^*}$ denotes the optimized patterns with physical constraints and $\operatorname{TV} $ refers to total variation which has been proven to be an effective image regularization method in previous studies \cite{40,41}. The parameter ${\lambda _3}$ is a weight factor and is set to ${10^{ - 10}}$. Through this iterative optimization process, Eq. 6 directs the second network ${R_\theta }$ to generate a 1D sequence $\tilde I$, as described in Eq. 1, which closely resembles the bucket signal $I$. The pseudocode for the training process and iterative optimization is provided in the appendix.

\section{Network structure}

\begin{figure}[htpb]
    \centering
    \includegraphics[width=1\linewidth]{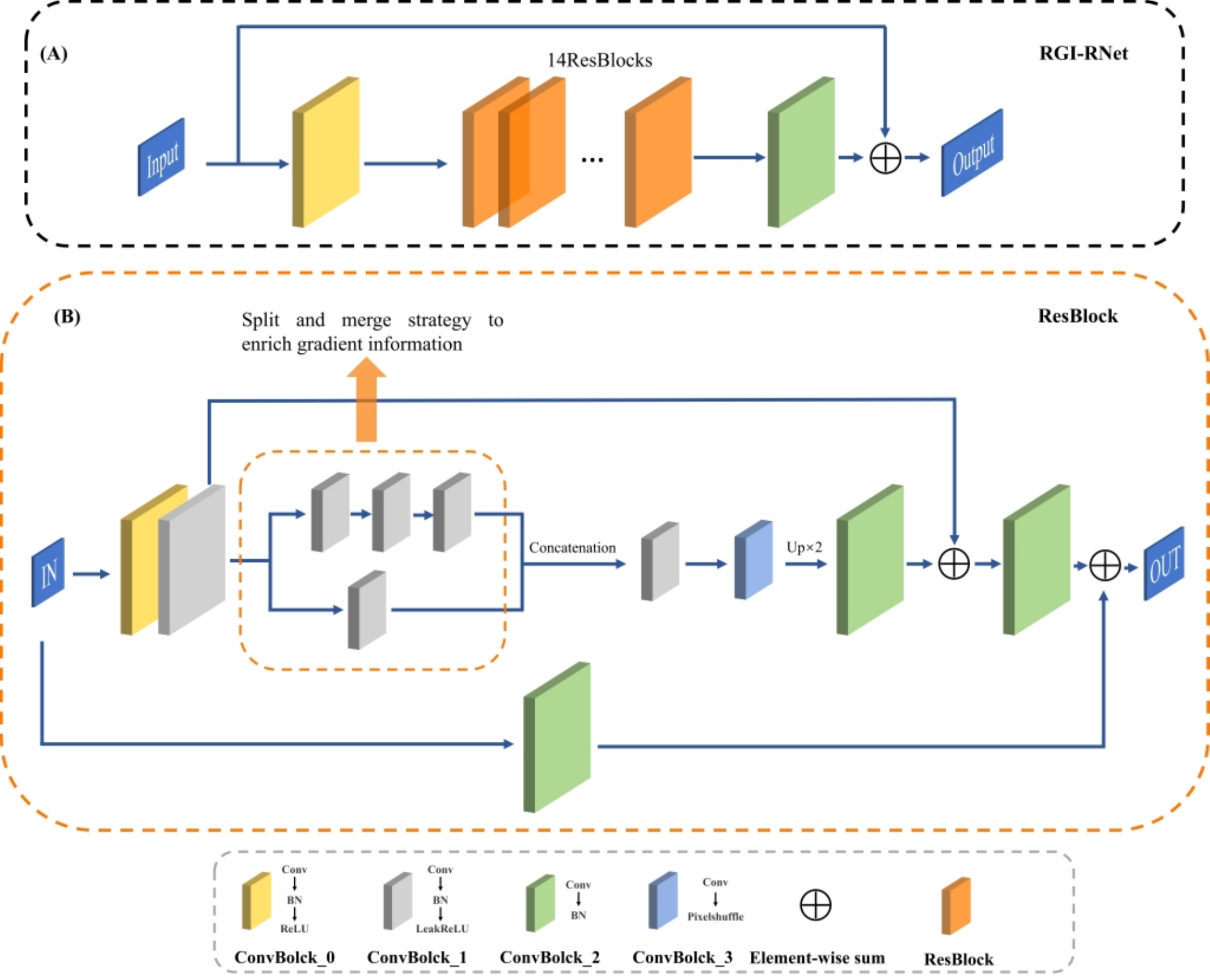}
    \caption{The framework of the RGI-RNet.}
    \label{fig:2}
\end{figure}

The structure of RGI-RNet shown in Fig. \ref{fig:2}(A) mainly consists of three parts. The first part is ConvBlock\_0, which serves as a shallow feature extraction block. The second part consists of multiple ResBlocks stacked in a chain-like manner to gradually refine the extracted features. The third part involves channel transformation using ConvBlock\_2. The DGI reconstruction, when used as input data, contains less effective information under compressed sampling. Additionally, the original data is more likely to be lost as the network depth increases, according to the information bottleneck principle \cite{44}. This makes it challenging to improve the network's reconstruction quality when using DGI as input data by simply increasing the model's depth. Inspired by CSPNet \cite{43}, we designed the residual blocks (ResBlocks) shown in Fig. \ref{fig:2}(B). Firstly, shallow feature extraction of the input data is achieved by stacking ConvBlock\_0 and ConvBlock\_1. Subsequently, the shallow feature layer is split into two parts through the channel, and the feature space size is downsampled to enhance inference speed. By employing this split and merge strategy across stages, we increase gradient paths and enrich gradient information. Furthermore, to preserve both input and shallow feature information, two skip connections are used. These methods enhance the network's learning capacity and improve its convergence. The structure details of each layer for RGI-RNet can be found in Fig. S3.

\section{Settings and Details}

The hyperparameter settings for the SPI-DNIO method proposed in this work can be divided into two parts. During the training process, we utilize the Adam optimizer to update the network’s parameters with constant learning rate $l{r_0} = {10^{ - 4}}$ and apply random flipping and brightness jitter for data augmentation. The other parameters of the optimizer adopt default values. All PT-Nets mentioned in this work are trained using a train-from-scratch strategy on the same device (13th Gen Intel(R) Core(TM) i9-13900K, NVIDIA GeForce RTX 4090) and converge within 1000 epochs. The training dataset contains 10000 images ($64 \times 64$ pixels in size) randomly selected from LFW, with the remaining images used as the validation dataset. For iterative optimization, the number of iterations is set to 400. The Adam optimizer with cosine annealing learning rate adjustment is exploited to update the DIP-Net’s parameters on the device (12th Gen Intel(R) Core(TM) i5-12500H, NVIDIA GeForce RTX 3050 Ti). The learning rate is adjusted with $l{r_{\max }} = {0.5^{2.2}} \times {10^{ - 4}}$, $l{r_{\min }} = {0.5^{6.6}} \times {10^{ - 4}}$, and ${T_{\max }}{\text{ =  iterations}}$ The weight decay is equal to 0.1, and other parameters are set to default values. For the ES strategy, the iteration process stops once the best SSIM has not improved and the number of iterations exceeds the patience threshold. The patience threshold is set to 125. In addition, the details of Unet are consistent with the reference \cite{33}.

\section{Results and Discussion}

\subsection{Simulations}
To demonstrate the effectiveness of SPI-DNIO in improving image reconstruction quality, we present a comparative analysis with PT-Net and DIP-Net. These networks share the same architecture, employing split and merge strategies across stages to enrich gradient information \cite{43}. The corresponding details are presented in Fig. \ref{fig:2}. In our comparison the DGI with the optimized patterns serves as a benchmark for evaluating the quality of image reconstruction. The target image, selected from the LFW validation dataset, is labeled as “Abdullah\_Gul\_0015”. The size of the recovered image is 64 $\times$ 64 and the sampling ratio is 25\%.

\begin{figure}[htpb]
    \centering
    \includegraphics[width=\linewidth]{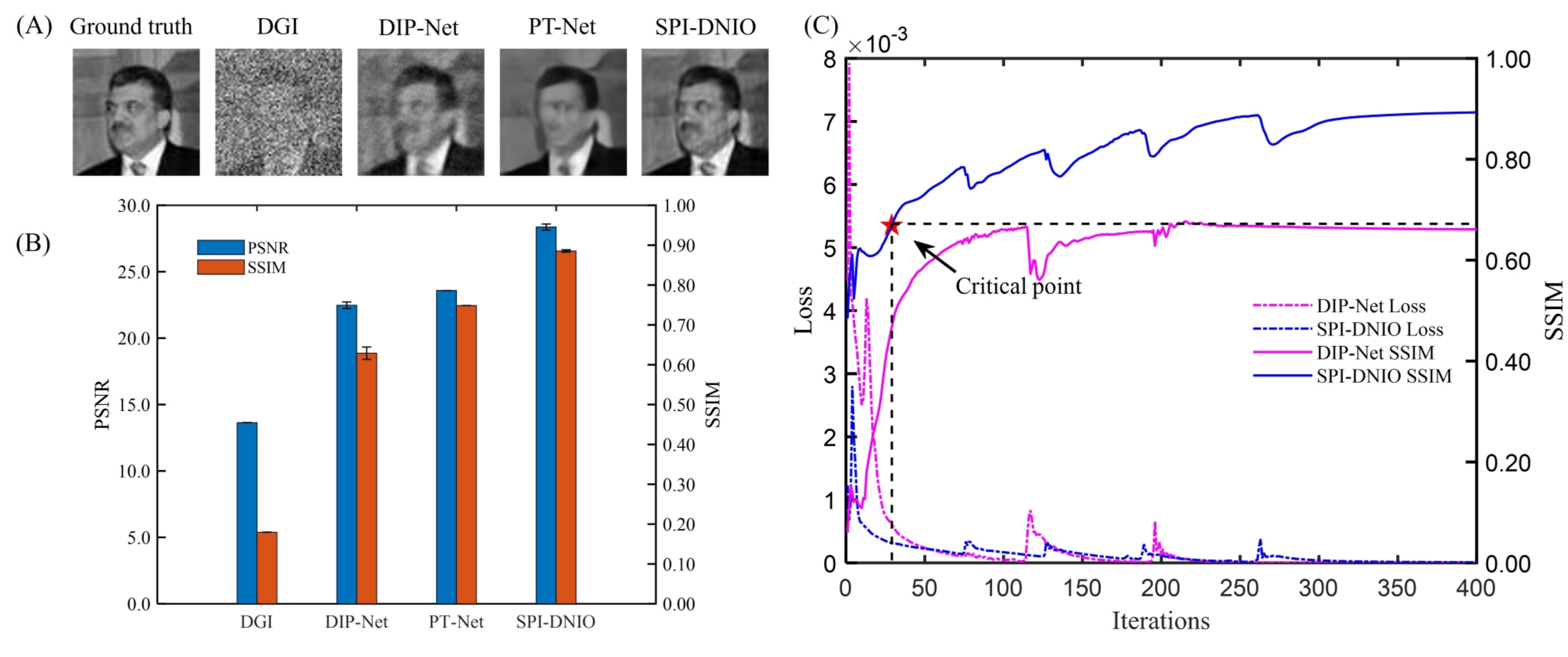}
    \caption{(A) The imaging results of the reconstruction networks including SPI-DNIO, DIP-Net, and the PT-Net for 64$\times$64 images at a 25\% sampling ratio. (B) Evaluation metrics for three reconstruction models and their corresponding error bars which are calculated through repeated experiments on the same device. The iteration step of SPI-DNIO and DIP-Net is 400. (C) The loss is calculated by Eq. (\ref{eq:6}) and the SSIM between the ground truth and the reconstruction along with the number of iteration steps from 0 to 400 for the target 'Abdullah\_Gul\_0015' in the LFW\cite{34} validation dataset. }
    \label{fig:3}
\end{figure}

As depicted in Fig. \ref{fig:3}(A), the PT-Net exhibits a smoother image reconstruction, which has a restoration capacity superior to that of the DIP-Net. However, its excessive restoring ability, which leads to facial information loss, also results in weaker fidelity compared to DIP-Net. The explanation for these phenomena is that the PT-Net parameters are updated based on the loss reduction for each image in the training dataset, resulting in the image reconstruction capability of PT-Net reflecting an averaged result. During the image reconstruction process of DIP-Net, it utilizes the bucket detection values of a single target image to update its parameters, effectively preserving the target image information. Therefore, the SPI-DNIO approach, which combines the restoring advantages of PT-Net with the ability of DIP-Net to reconstruct high-fidelity images, is effective in improving image reconstruction quality and reducing the number of iterative optimizations compared to some previous results \cite{31,33}. For the critical point marked in Fig. \ref{fig:3}(C), we observe that the structural similarity index (SSIM) of the SPI-DNIO reconstruction result exceeds that of DIP-Net with fewer optimization iterations. This phenomenon demonstrates that the efficiency of SPI-DNIO is improved by achieving higher performance with fewer iterations. The detailed quantitative evaluation indices (SSIM and peak signal-to-noise ratio, PSNR) are displayed in Fig. \ref{fig:3}(B), indicating that the performance of SPI-DNIO is significantly higher than that of DGI reconstruction and also superior to the image reconstruction results of PT-Net and DIP-Net alone. In addition, we consider the results of using loss ${\mathcal{L}_{all\_2}}$ to fine-tune PT-Net. The corresponding comparison results are presented in Fig. S1. Compared to the fine-tuning PT-Net, the SPI-DNIO effectively improves the image reconstruction ability with the same number of iterations. Moreover, a decrease in fine-tuning PT-Net performance can occur in our test if the fine-tuning learning rate is set improperly. 

Subsequently, to comprehensively investigate the robustness of the SPI-DNIO, we consider not only the reconstruction performance of this method under different noise levels but also the scenario where the target object is inconsistent with the training dataset. The target objects are selected from the LFW validation dataset and the DIV2K validation dataset, respectively. For an imaging system, accurately describing the noise distribution in experiments is challenging. Therefore, we simplify the noise modeling of the system by only considering additive Gaussian noise in the bucket detector. The bucket signal with noise can be expressed as 
\begin{align}
 \begin{gathered}
{I_{m}} = \int {{H_m}({x_i},{y_i})T({x_i},{y_i})dxdy}  + n(\mu ,\sigma ),
  \end{gathered}
 \label{eq:7}
\end{align}
where $n(\mu ,\sigma )$ is the additive noise possesses $\mu$ mean Gaussian distribution with a standard deviation of $\sigma$. The $\mu$ we set to zero. Thus, the detection SNR (dSNR) can be defined as follows 
\begin{align}
 \begin{gathered}
{\text{dSNR = 10lo}}{{\text{g}}_{10}}\frac{{\left\langle I \right\rangle }}{\sigma }
  \end{gathered}
 \label{eq:8}
\end{align}
to describe the noise level of the detected signal.

\begin{figure}[htpb]
    \centering
    \includegraphics[width=\linewidth]{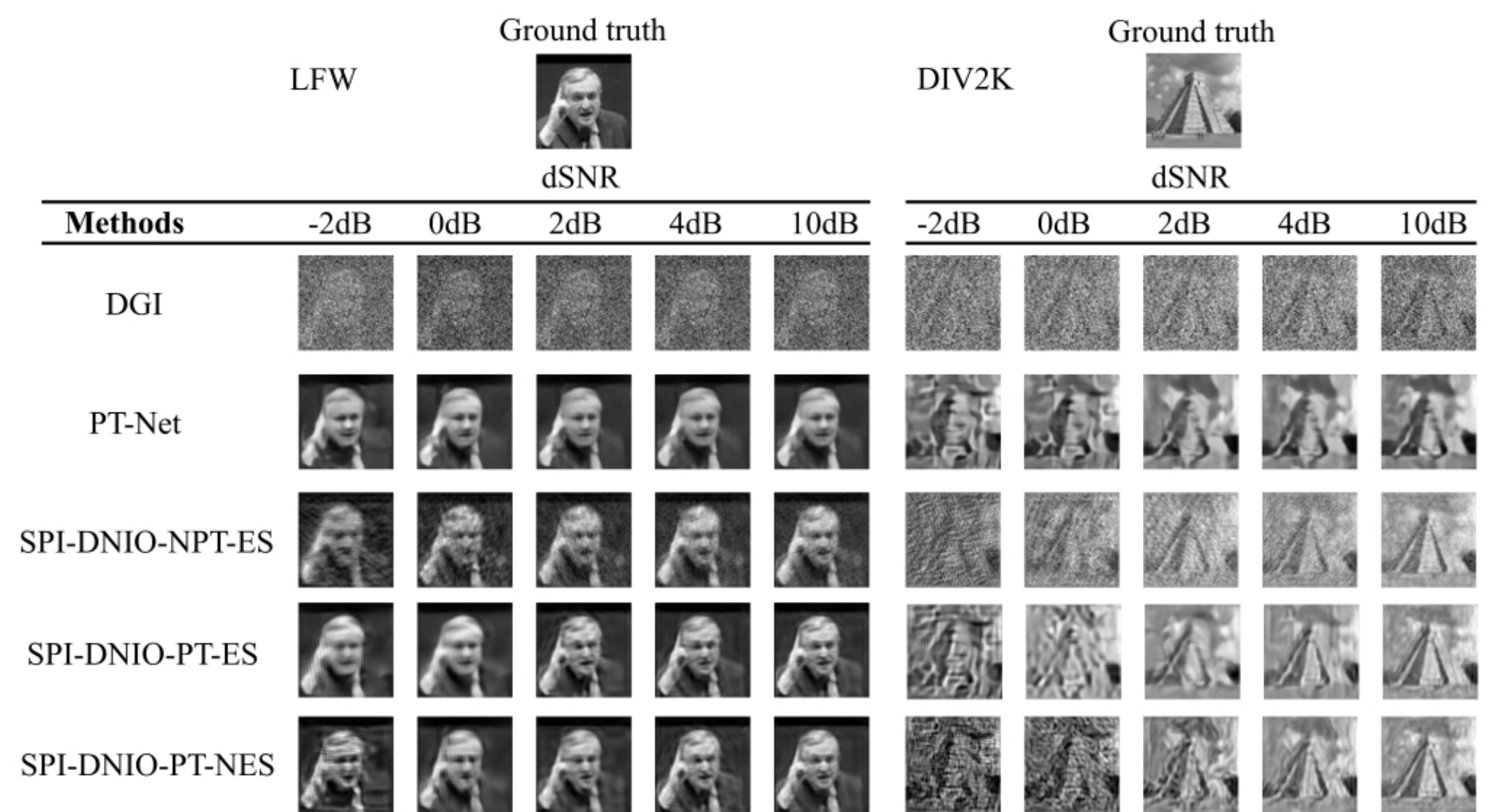}
    \caption{Robustness comparisons of different SPI reconstruction methods under various dSNR levels. The method naming explanations are as follows. PT represents the first frozen RGI-RNet in SPI-DNIO, which loads the PT-weight, while NPT stands for the random initialization of the first frozen network. ES refers to adopting the ES strategy during iterative optimization while NES means not employing the ES strategy. The image size is 64 $\times$ 64 and the sampling ratio is set to 25\%. The "Face" and the "Pyramid" object images are selected from the LFW\cite{34} and DIV2K validation datasets, respectively. The iteration step of the SPI-DNIO is 400.}
    \label{fig:4}
\end{figure}

Meanwhile, multiple variants of SPI-DNIO are designed for comparative analysis. The method naming descriptions are as follows. PT-Net stands for the RGI-RNet trained in the first step. PT represents the first frozen RGI-RNet in SPI-DNIO, which loads the PT-weight, while NPT stands for the random initialization of the first frozen network. Early stopping (ES) refers to adopting the ES strategy during iterative optimization while NES means not employing the ES strategy. The target image size is 64 $\times$ 64 and the sampling ratio is set to 25\%.

Figure \ref{fig:4} exhibits the visualization results of different SPI reconstruction methods at different noise levels. From Fig. \ref{fig:4} we can observe that the face image is relatively well reconstructed by SPI-DNIO-PT-ES even at a dSNR as low as -2 dB while the main part of the reconstructed pyramid is obscured by false textures, with only the edges visible. By comparing the third row with the fourth row, it is evident that the denoising capability of SPI-DNIO is enhanced by the data priors from PT, rather than simply stacking two networks. Besides, utilizing the ES strategy clearly improves image reconstruction quality at low dSNR, while the advantage of the ES strategy diminishes as the dSNR increases. The explanation for this phenomenon is that the DIP-Net first learns the desired visual content and then picks up potential modeling and observational noise during the learning process \cite{42}. In the case of the SPI-DNIO method, the second network (DIP-Net) plays a crucial role in the iterative optimization. Extremely low dSNR causes the DIP-Net to overfit the noise distribution, resulting in false textures and a decrease in the reconstructed image quality. Therefore, adopting the ES strategy can prevent this phenomenon occurring. To give a quantitative evaluation of the noise immunity of these methods, the PSNR and SSIM are shown in Table \ref{tab:1}. From Table \ref{tab:1}, we can notice that the SPI-DNIO-PT-ES exhibits almost the best reconstruction performance in a noisy environment, compared with other SPI-DNIO variants. Meanwhile, the data prior from PT-Net improves target reconstruction, whether the target is similar to the training dataset or not. However, the performance of PT-Net for reconstructing the pyramid is better than SPI-DNIO-PT-ES at a dSNR of -2 dB. The explanation for this phenomenon is that PT-Net processing a target that is significantly different from the training dataset generates artificial textures, inducing noise distribution overfitting in DIP-Net. Therefore, degrading SPI-DNIO's reconstruction performance at -2 dB dSNR.
\begin{table}[htpb]
    \caption{The metrics of different SPI reconstruction methods on the SSIM and the PSNR at different noise levels. The highest values of the SSIM or PSNR are in bold. The abbreviations in the method possess the same meaning as the description in Fig. \ref{fig:4}.}
    \label{tab:1}
    \centering
    \resizebox{\textwidth}{!}{ 
    \begin{tabular}{lcccccc}
        \toprule
        \multicolumn{2}{l}{} & \multicolumn{5}{c}{dSNR} \\
        \cmidrule{3-7}
        \multicolumn{2}{c}{} & -2dB & 0dB & 2dB & 4dB & 10dB \\
        \cmidrule{1-7}
        \multirow{1}{*}{Object} & \multirow{1}{*}{Methods} & PSNR/SSIM & PSNR/SSIM & PSNR/SSIM & PSNR/SSIM & PSNR/SSIM \\
        
        \midrule
        \multirow{5}{*}{"Face"} & DGI & 12.71/0.1625 & 12.71/0.1679 & 12.68/0.1701 & 12.66/0.1708 & 12.61/0.1710 \\
        & PT-Net & 21.95/0.7114 & 22.88/0.7370 & 22.71/0.7621 & 22.69/0.7694 & 23.35/0.7792\\
        & SPI-DNIO-NPT-ES & 18.2/0.5204 & 20.12/0.5822 & 21.51/0.6483 & 22.78/0.6829 & 23.67/0.7301 \\  
        & SPI-DNIO-PT-ES & \textbf{22.23}/\textbf{0.7279} & \textbf{23.67}/\textbf{0.7773} & \textbf{24.53}/\textbf{0.8227} & 24.99/\textbf{0.8504} & \textbf{27.72}/0.882 \\
        & SPI-DNIO-PT-NES & 19.77/0.6155 & 23.3/0.766 & 24.21/0.8149 & \textbf{25.73}/0.8455 & 27.05/\textbf{0.8845} \\
        \midrule
        \multirow{5}{*}{"Pyramid"} & DGI & 14.44/0.1662 & 15.25/0.2094 & 15.81/0.2426 & 15.93/0.2634 & 15.43/0.2787 \\
        & PT-Net & \textbf{16.21}/\textbf{0.3327} & 17.04/\textbf{0.42} & 19.04/0.5238 & 19.57/0.5732 & 19.6/0.6114\\
        & SPI-DNIO-NPT-ES & 14.22/0.1954 & 17.42/0.2852 & 18.19/0.3822 & 19.65/0.4497 & 20.04/0.6891 \\
        & SPI-DNIO-PT-ES & 15.39/0.3028 & \textbf{17.58}/0.4164 & \textbf{21.16}/\textbf{0.5806} & \textbf{22.09/0.6287} & 22.25/\textbf{0.7501} \\
        & SPI-DNIO-PT-NES & 11.75/0.2268 & 13.77/0.3312 & 19.49/0.4843 & 21.93/0.6221 & \textbf{22.76}/0.7494 \\
        \bottomrule
    \end{tabular}
    }

\end{table}

\begin{figure}[htpb]
    \centering
    \includegraphics[width=\linewidth]{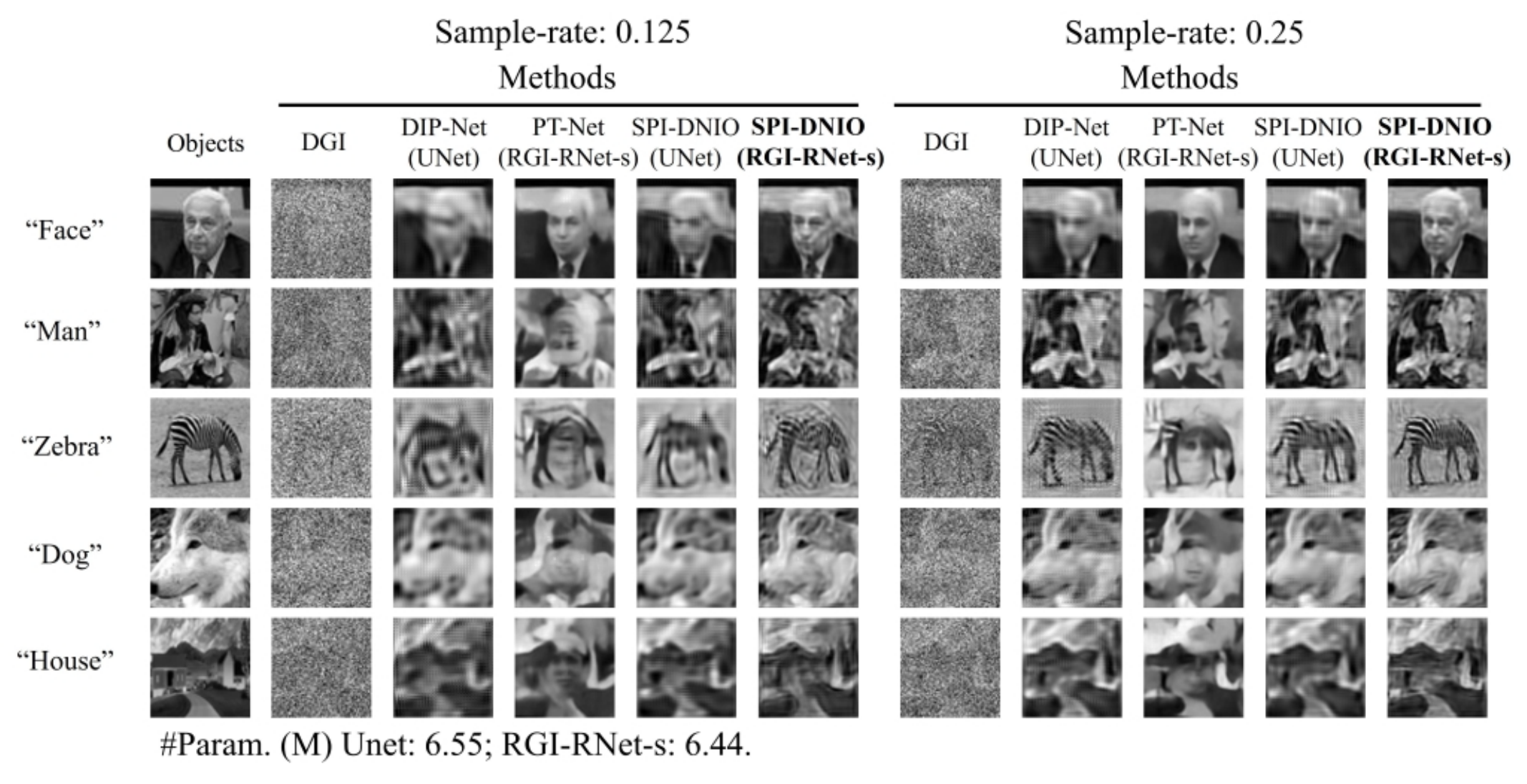}
    \caption{Comparison of different SPI reconstruction approaches at different sampling rates. The target objects are selected from diverse datasets including LFW, SET14, DIV2K, and LSDIR. The RGI-RNet-s is a scaling down of the RGI-RNet, only decreasing the number of ResBlock to 9 and the structure of Unet is consistent with the reference \cite{33}. The image size is $64 \times 64$. The iteration step for SPI-DNIO and DIP-Net is 400. The networks utilized are indicated in parentheses. The face images were taken from LFW\cite{34}}
    \label{fig:5}
\end{figure}

To verify the generalization performance of SPI-DNIO and assess how its components, PT-Net and DIP-Net, perform across a wide range of data types, target objects are selected from different datasets (LFW, SET14, DIV2K, and LSDIR), which include "Face", "Humans", "Animals", and "Buildings". Meanwhile, various SPI reconstruction methods are used for comparison and the content in parentheses refers to the specific network structure utilized. RGI-RNet-s represents a scaling version of RGI-RNet, with parameters approximately equal to those of Unet. The optimized patterns of DIP-Net (Unet) and PT-Net (RGI-RNet-s) are consistent with SPI-DNIO (Unet) and SPI-DNIO (RGI-RNet-s), respectively and the patterns of DGI are the same as PT-Net (RGI-RNet-s).

\begin{table}[H]
    \caption{The quantitative comparison of different SPI reconstruction approaches on the SSIM and the PSNR at different sampling rates. The highest values of the SSIM or PSNR are in bold.}
    \label{tab:2}
    \centering
    \resizebox{\textwidth}{!}{
    \begin{tabular}{cccccccccccc}
        \toprule
         &  \multicolumn{5}{c}{Sample-rate: 0.125} & & \multicolumn{5}{c}{Sample-rate: 0.25} \\
        \cmidrule(lr){2-6} \cmidrule(lr){8-12}
        
        \multirow{2}{*}{Object\textbackslash Methods} & \multirow{2}{*}{DGI} &DIP-Net  & PT-Net & SPI-DNIO & \textbf{SPI-DNIO} & &  \multirow{2}{*}{DGI}  & DIP-Net & PT-Net & SPI-DNIO & \textbf{SPI-DNIO} \\
        & &(Unet)  & (RGI-RNet-s) & (Unet) & \textbf{(RGI-RNet-s)} & &  &(Unet)  & (RGI-RNet-s) & (Unet) & \textbf{(RGI-RNet-s)} \\
        
            \midrule    
            \multirow{1}{*}{} & PSNR/SSIM & PSNR/SSIM & PSNR/SSIM & PSNR/SSIM & PSNR/SSIM &  & PSNR/SSIM & PSNR/SSIM & PSNR/SSIM & PSNR/SSIM& PSNR/SSIM \\
            \midrule 
             \multirow{2}{*}{"Face"} & \multirow{2}{*}{10.15/0.1048} & \multirow{2}{*}{21.81/0.6031} & \multirow{2}{*}{22.73/0.7042} & \multirow{2}{*}{23.71/0.6885} & \multirow{2}{*}{\textbf{24.61}/\textbf{0.7788}} & \multirow{2}{*}{} & \multirow{2}{*}{10.85/0.1408} & \multirow{2}{*}{24.94/0.7295} & \multirow{2}{*}{25.44/0.7769} & \multirow{2}{*}{26.51/0.7836} & \multirow{2}{*}{\textbf{29.22}/\textbf{0.8908}} \\\\
        \multirow{2}{*}{"Man"}    & \multirow{2}{*}{12.20/0.1807} & \multirow{2}{*}{18.05/0.4808} & \multirow{2}{*}{13.10/0.3759} & \multirow{2}{*}{18.67/0.5424} & \multirow{2}{*}{\textbf{18.83/0.5808}} &\multirow{2}{*}{} & \multirow{2}{*}{12.16/0.2459} & \multirow{2}{*}{18.58/0.6386} & \multirow{2}{*}{18.10/0.5728} & \multirow{2}{*}{20.75/0.7229} & \multirow{2}{*}{\textbf{22.43}/\textbf{0.7848}} \\ \\
        \multirow{2}{*}{"Zebra"}   & \multirow{2}{*}{14.14/0.2406} & \multirow{2}{*}{16.35/0.3262} & \multirow{2}{*}{16.79/0.4243} & \multirow{2}{*}{16.58/0.4292} & \multirow{2}{*}{\textbf{19.01}/\textbf{0.5334}} &\multirow{2}{*}{} & \multirow{2}{*}{15.03/0.2847} & \multirow{2}{*}{19.35/0.5234} & \multirow{2}{*}{14.76/0.4847} & \multirow{2}{*}{17.85/0.5606} & \multirow{2}{*}{\textbf{22.89}/\textbf{0.7289}} \\  \\
        \multirow{2}{*}{"Dog"}& \multirow{2}{*}{12.30/0.1421} & \multirow{2}{*}{19.2/0.5922} & \multirow{2}{*}{15.63/0.4996} & \multirow{2}{*}{19.14/\textbf{0.6571}} & \multirow{2}{*}{\textbf{20.16}/0.6506} &\multirow{2}{*}{} & \multirow{2}{*}{12.96/0.2038} & \multirow{2}{*}{19.02/0.6555} & \multirow{2}{*}{16.67/0.5776} & \multirow{2}{*}{19.45/0.6923} & \multirow{2}{*}{\textbf{21.46}/\textbf{0.7561}} \\  \\
        \multirow{2}{*}{"House"} & \multirow{2}{*}{12.59/0.1494} & \multirow{2}{*}{19.83/0.4648} & \multirow{2}{*}{19.19/0.5226} & \multirow{2}{*}{21.34/0.5805} & \multirow{2}{*}{\textbf{21.92}/\textbf{0.6268}} &\multirow{2}{*}{} & \multirow{2}{*}{12.72/0.2129} & \multirow{2}{*}{20.82/0.6754} & \multirow{2}{*}{17.69/0.5815} & \multirow{2}{*}{22.34/0.7022} & \multirow{2}{*}{\textbf{24.12}/\textbf{0.7772}} \\
        \\
        \bottomrule
    \end{tabular}
    }

\end{table}

In the case of single network reconstruction, shown in Figure \ref{fig:5}, we can notice that the PT-Net can reconstruct the object only when the target image is similar to the training dataset. Otherwise, false faces (false textures are related to the type of training dataset) will appear in PT-Net reconstruction, and this phenomenon becomes more pronounced as the sampling rate decreases. However, DIP-Net can reconstruct various data types, unlike PT-Net. Adopting SPI-DNIO reconstruction can overcome the limitations imposed by the data prior of PT-Net and deliver better reconstruction results than DIP-Net, even when the target objects are not similar to the training dataset. At a sampling rate of 0.125, SPI-DNIO (RGI-RNet-s) is still able to reconstruct the approximate shape of the objects, demonstrating the superiority of the RGI-RNet-s design, which enhances the information complexity of gradient flow to improve the learning ability for input objects with lower sampling rates. For the DIP-Net (Unet) and SPI-DNIO (Unet) results shown in Fig. \ref{fig:5}, we can observe grid-like artifacts appear in reconstruction due to the existence of transpose convolution \cite{45}. Furthermore, the quantitative evaluations (PSNR and SSIM) are shown in Table \ref{tab:2} and we can find that the SPI-DNIO (RGI-RNet-s) almost achieved the best SSIM and PSNR among the methods studied.

\subsection{Experiments}

\begin{figure}[htbp]
    \centering
    \includegraphics[width=0.9\linewidth]{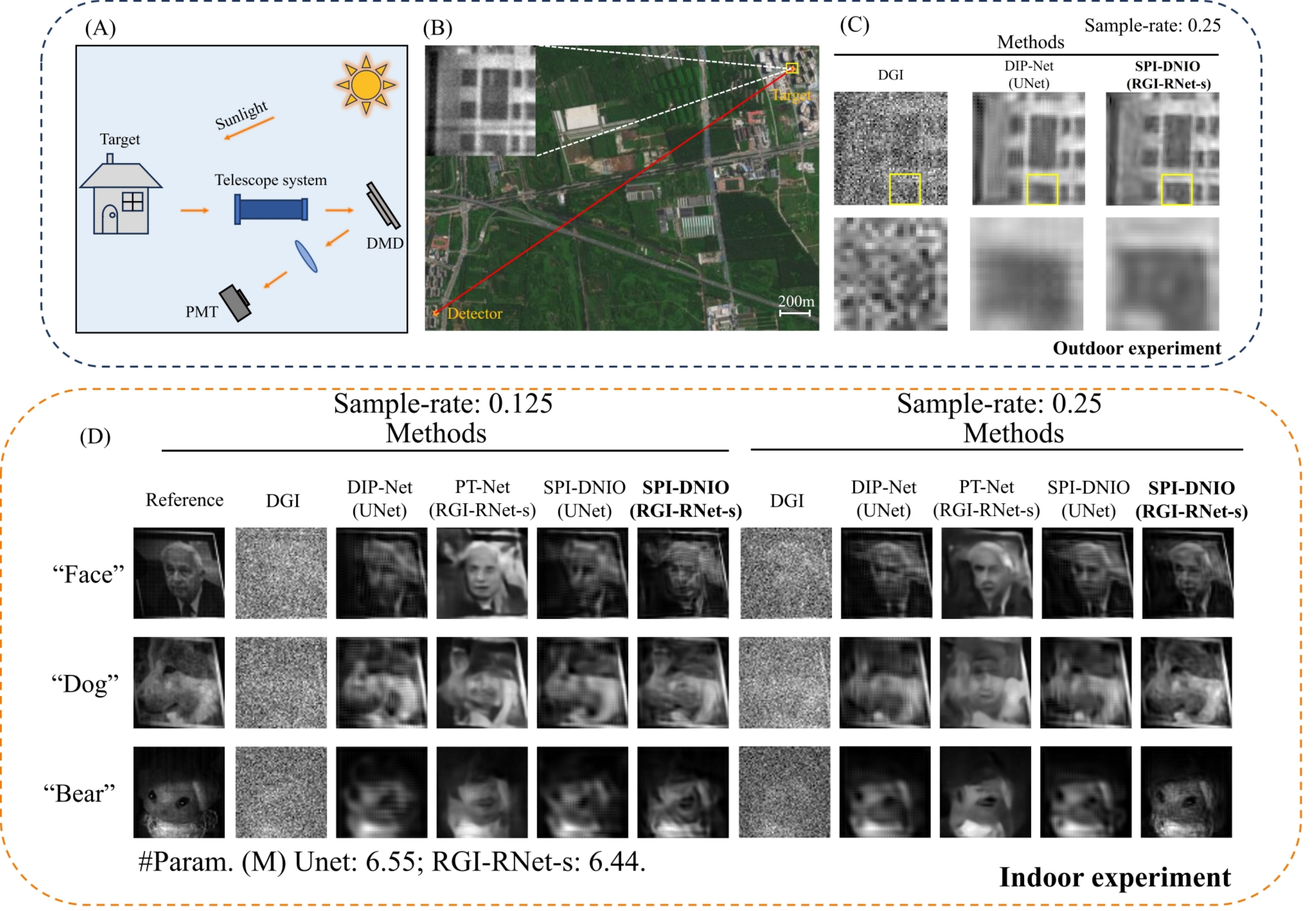}
    \caption{Experimental results of different SPI reconstruction approaches at various sampling rates. (A) Schematic diagram of the outdoor experiment. DMD: digital micromirror device; PMT: photomultiplier tube. (B) Satellite image of the outdoor experiment scenario. The inset in the top left shows the target image, which is reconstructed by DGI using cake-cutting Hadamard basis sort patterns at full sampling \cite{14}. (C) Outdoor experimental results. The first row shows the reconstruction results utilizing different SPI reconstruction approaches and the second row displays the locally enlarged result of the yellow box. (D) Indoor experimental results. The image size is $64 \times 64$. The iteration step for SPI-DNIO and DIP-Net is 400. The parentheses indicate the network utilized. }
    \label{fig:6}
\end{figure}

To verify the reconstruction performance and robustness of SPI-DNIO in real environments, we apply it to both the indoor experiment with active lighting and the outdoor long-range experiment with passive lighting. For the indoor experiments, different types of objects are used for imaging. The "Face" and "Dog" images, which are consistent with those shown in Fig. \ref{fig:5}, are printed on white paper, while the "Bear" is a physical toy. To obtain the reference reconstruction results as the ground truth, we exploit DGI reconstruction with cake-cutting Hadamard basis sort patterns at full sampling. The corresponding experimental setups are exhibited in the supplementary document Fig. S2. From Fig. \ref{fig:6}(D), we can find that the SPI-DNIO (RGI-RNet-s) successfully reconstructs different targets in a real environment, demonstrating the effectiveness of the SPI-DNIO framework and RGI-RNet-s structure compared to other methods in the study. The detailed analysis for the component of SPI-DNIO is provided in the discussion in Fig. \ref{fig:5}. The quantitative evaluations for Fig. \ref{fig:6}(D) are shown in Table \ref{tab:3}. Furthermore, we apply the proposed method to the outdoor experiment. We select a building located about 2.9 km away from our detector as the imaging object. From the enlarged view shown in Fig. \ref{fig:6}(C), we observe that the proposed method can reconstruct more edge details. In our experiment, the SPI-DNIO requires 18.1s for 400 iterations, which is recorded by running SPI-DNIO on the same device (NVIDIA GeForce RTX 3050 Ti) repeatedly 10 times and collecting the reconstruction time with the tqdm Library. The performance of SPI-DNIO approximates an 80\% convergence value with 100 iterations, similar to Fig. \ref{fig:3}(C), and the reconstruction time is also related to devices.

\begin{table}[H]
    \caption{The quantitative evaluations of different SPI approaches for indoor experimental results on the SSIM and the PSNR at various sampling rates. The highest values of the SSIM or PSNR are in bold.}
    \label{tab:3}
    \centering
    \resizebox{\textwidth}{!}{
    \begin{tabular}{cccccccccccc}
        \toprule
         &  \multicolumn{5}{c}{Sample-rate: 0.125} & & \multicolumn{5}{c}{Sample-rate: 0.25} \\
        \cmidrule(lr){2-6} \cmidrule(lr){8-12}
        
        \multirow{2}{*}{Object\textbackslash Methods} & \multirow{2}{*}{DGI} &DIP-Net  & PT-Net & SPI-DNIO & \textbf{SPI-DNIO} & &  \multirow{2}{*}{DGI}  & DIP-Net & PT-Net & SPI-DNIO & \textbf{SPI-DNIO} \\
        & &(Unet)  & (RGI-RNet-s) & (Unet) & \textbf{(RGI-RNet-s)} & &  &(Unet)  & (RGI-RNet-s) & (Unet) & \textbf{(RGI-RNet-s)} \\
        \midrule    
        \multirow{1}{*}{} & PSNR/SSIM & PSNR/SSIM & PSNR/SSIM & PSNR/SSIM & PSNR/SSIM &  & PSNR/SSIM & PSNR/SSIM & PSNR/SSIM & PSNR/SSIM& PSNR/SSIM \\
        \midrule
         \multirow{2}{*}{"Face"} & \multirow{2}{*}{7.71/0.0909} & \multirow{2}{*}{23.49/0.5802} & \multirow{2}{*}{13.37/0.4287} & \multirow{2}{*}{24.01/0.6514} & \multirow{2}{*}{\textbf{24.68}/\textbf{0.6627}} & \multirow{2}{*}{} & \multirow{2}{*}{9.06/0.1245} & \multirow{2}{*}{25.93/0.6834} & \multirow{2}{*}{18.32/0.6241} & \multirow{2}{*}{26.37/0.7056} & \multirow{2}{*}{\textbf{28.81}/\textbf{0.7971}} \\
         \\
        \multirow{2}{*}{"Dog"}    & \multirow{2}{*}{10.86/0.1036} & \multirow{2}{*}{20.93/0.5555} & \multirow{2}{*}{20.03/0.5452} & \multirow{2}{*}{23.22/0.6076} & \multirow{2}{*}{\textbf{25.18}/\textbf{0.699}} &\multirow{2}{*}{} & \multirow{2}{*}{10.45/0.1336} & \multirow{2}{*}{25.07/0.6692} & \multirow{2}{*}{20.47/0.6004} & \multirow{2}{*}{\textbf{25.8}/0.7132} & \multirow{2}{*}{24.6/\textbf{0.7991}} \\ \\
        \multirow{2}{*}{"Bear"}   & \multirow{2}{*}{8.55/0.0684} & \multirow{2}{*}{25.92/0.6663} & \multirow{2}{*}{20.96/0.6578} & \multirow{2}{*}{\textbf{26.28}/0.6816} & \multirow{2}{*}{26.11/\textbf{0.735}} &\multirow{2}{*}{} & \multirow{2}{*}{10.40/0.1083} & \multirow{2}{*}{27.32/0.7434} & \multirow{2}{*}{22.338/0.6881} & \multirow{2}{*}{26.59/0.751} & \multirow{2}{*}{\textbf{28.93}/\textbf{0.7827}} \\  \\
        \bottomrule
    \end{tabular}
    }
\end{table}

\section{Conclusion}

In summary, we demonstrate the effectiveness of the SPI-DNIO framework, which combines the advantages of DD-Net and DIP-Net, achieving high-quality reconstruction for various targets, and decreasing optimized iteration to hundreds of steps. Our quantitative evaluations indicate the superior performance of the SPI-DNIO framework compared with a single network reconstruction framework. In addition, the PT-Net as the first network of SPI-DNIO, effectively increases SPI-DNIO denoising capability regardless of whether the target image is similar to training set features or not. Even if the dSNR decreases to 2dB, the targets at a 25\% sampling ratio can be reconstructed by the SPI-DNIO. Furthermore, we study the generalization of SPI-DNIO by exploiting different types of targets from LFW \cite{34}, DIV2K \cite{35}, Set14 \cite{36}, and LSDIR \cite{37}, respectively. Both simulations and experimental evaluations demonstrate the robustness of our SPI-DNIO across diverse environments, spanning from actively lit indoor settings to passively illuminated outdoor scenarios. The results validate SPI-DNIO's exceptional performance in reconstruction accuracy, generalization capability, and robustness.

\section*{Disclosures} The authors declare no conflicts of interest.

\section*{Data Availability} Data and code used in this study are publicly available at \url{https://github.com/2020shijingyi/SPI-DNIO}.

\bibliographystyle{unsrt}  
\bibliography{sample}  

\clearpage
\renewcommand{\thefigure}{S\arabic{figure}}
\setcounter{figure}{0}

\section*{Supplementary materials}

\begin{figure}[H]
    \centering
    \includegraphics[width=\linewidth,height=0.5\textheight, keepaspectratio]{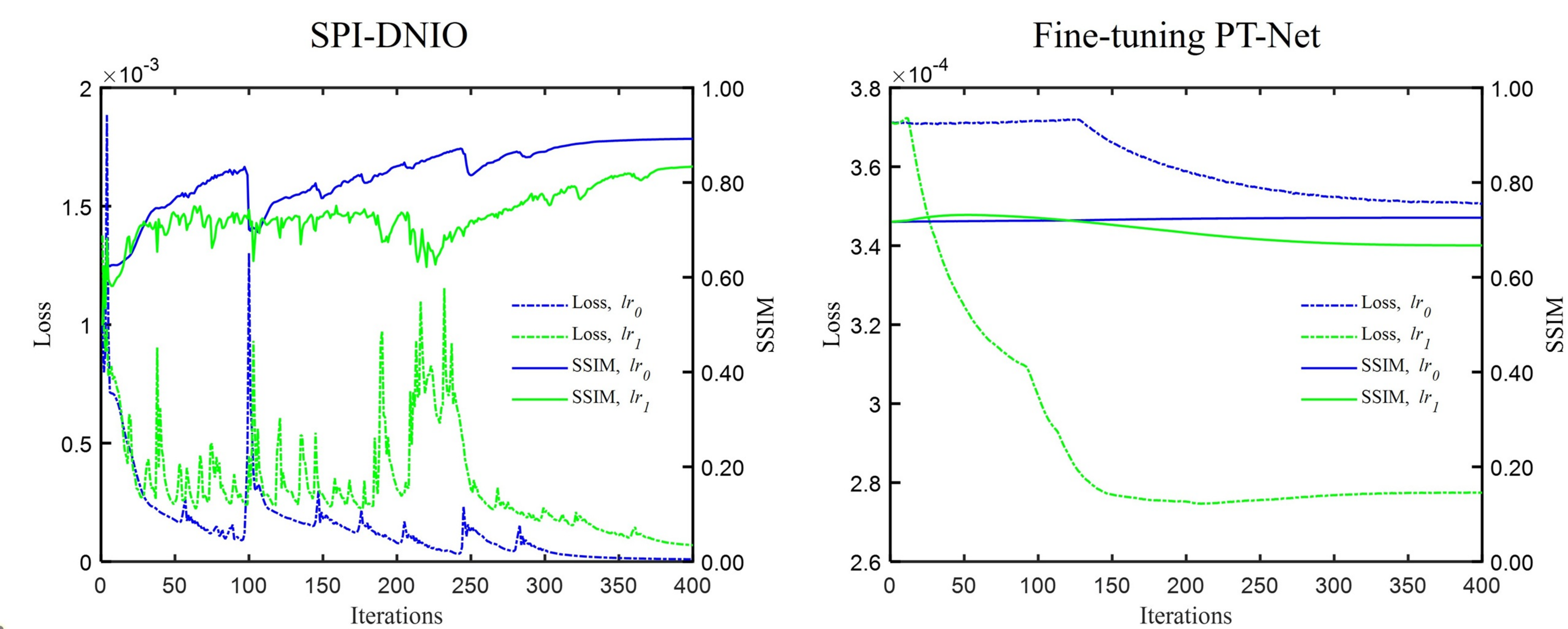}
    \caption{Quantitative comparison between SPI-DNIO and Fine-tuning PT-Net. The learning rates $l{r_0} = l{r_{\max }} = {0.5^{2.2}} \times {10^{ - 4}}$ and $l{r_1} = 10l{r_{\max }}$. The target image and other hyperparameter settings are consistent with Fig. 2.}
    \label{fig:S1}
\end{figure}

\begin{figure}[H]
    \centering
    \includegraphics[width=\linewidth,height=0.5\textheight, keepaspectratio]{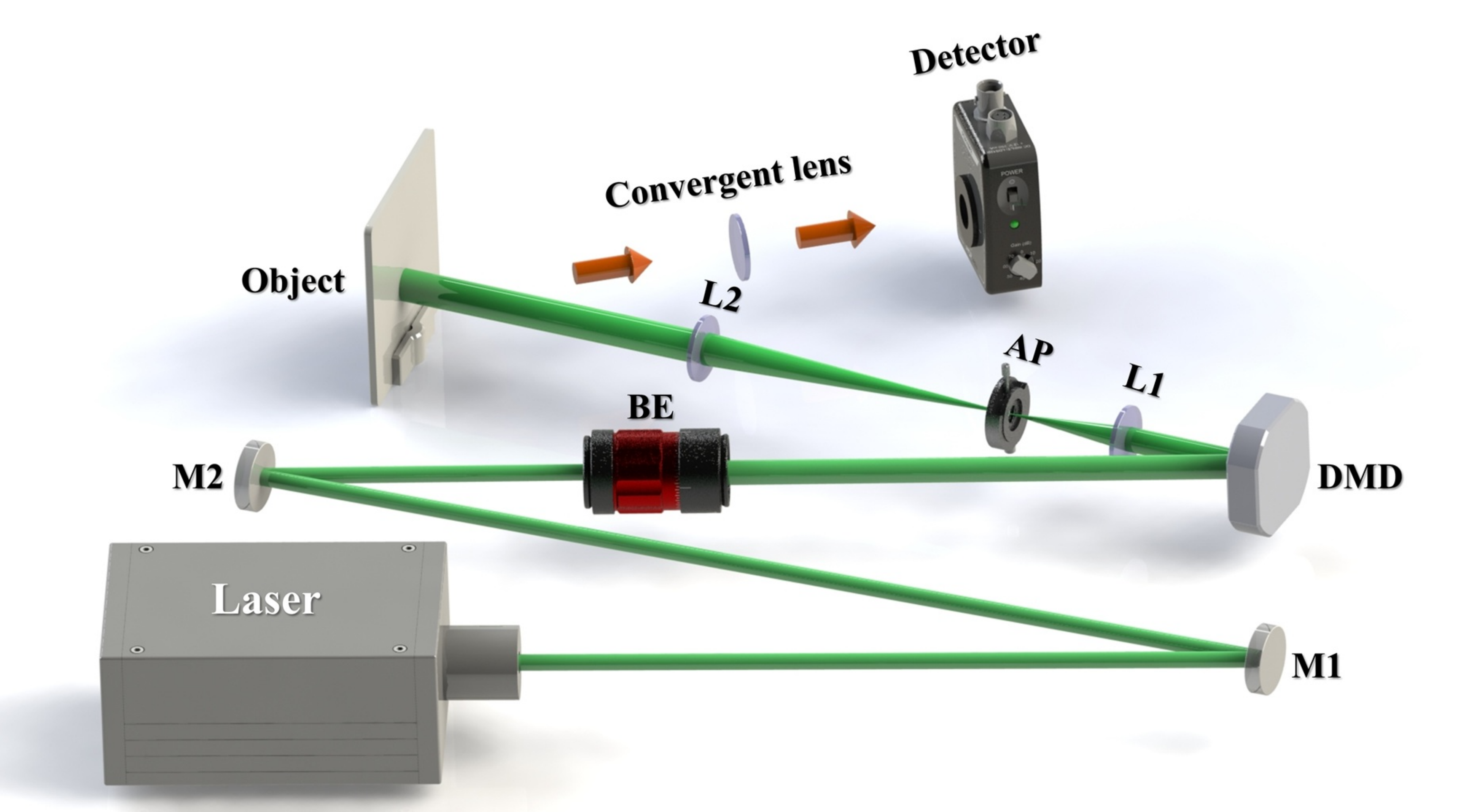}
    \caption{The experimental setup for indoor experiments. M1-M2: mirror; BE: beam expander; DMD: digital micromirror device; L1-L2: lens; AP: aperture.}
    \label{fig:S2}
\end{figure}

\begin{figure}[H]
    \centering
    \includegraphics[width=\linewidth, height=0.8\textheight, keepaspectratio]{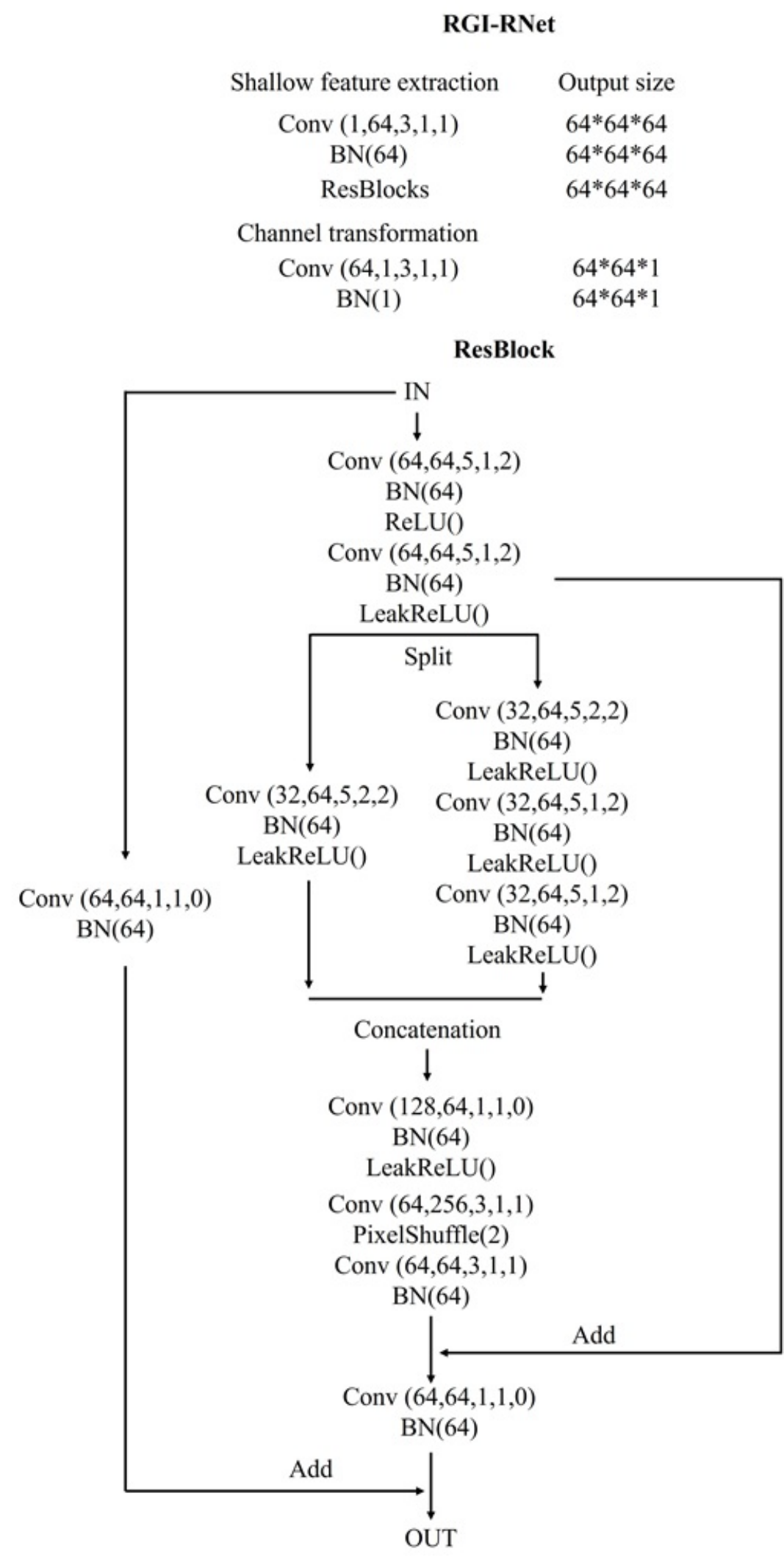}
    \caption{Details of the RGI-RNet. The input parameter of Conv refer to (channel\_in, channel\_out, kernel\_size, stride, padding).}
    \label{fig:S3}
\end{figure}

\begin{algorithm}[htpb]
\caption{Training process. $N=1000$, $\lambda _1=0.01$, $\lambda _2=1$, $l{r_0}={10^{ - 4}}$.}
\begin{algorithmic}[1]
\Statex \textbf{Input:} $T(h,w)$, $\mathrm{H}(h,w,M)$ and $R_\theta$
\Statex \textbf{Output:} ${\mathrm{H}}^*(h,w,M)$ and ${R_1}$
\State \textbf{initialize:} randomly initialize the parameters $\theta$ in the neural network $\mathcal{R}_\theta$ and the patterns $\mathrm{H}(h,w,M)$
\For{$epoch = 1, 2, ..., N$}
    \State ${I} \gets \sum\limits_{{h_i},{w_i}} {\mathrm{H}({h_i},{w_i})T({h_i},{w_i})}$
    \State $O_{DGI} \gets DGI(\mathrm{H}; I)$
    \State $\boldsymbol{\Phi} \gets \mathrm{reshape}(\mathrm{H})$
    \State ${{\cal L}_{{\rm{H}}}} \gets {\lambda _1}{\sum\limits_{ij} {(1 - {\boldsymbol{\Phi} _{ij}})} ^2}{(1 + {\boldsymbol{\Phi} _{ij}})^2} + {\lambda _2}{\left| {{\rm \boldsymbol{I_M}} - \frac{1}{M}{\boldsymbol{\Phi} ^T}\boldsymbol{\Phi} } \right|^2}$
    \State ${{\cal L}_{\theta }} \gets {\mathcal{L}_1}({R_\theta }({O_{DGI}}),T)$
    \State ${{\cal L}_{\theta,\mathrm{H}}} \gets{{\cal L}_{\theta }}+{{\cal L}_{{\rm{H}}}}$
    \State $\theta,\mathrm{H} \gets \text{Adam}(\nabla_{\theta,\mathrm{H}} \mathcal{L}_{\theta,\mathrm{H}}, l{r_0})$
\EndFor
\end{algorithmic}

\end{algorithm}

\begin{algorithm}[H]
\caption{Iterative optimization. $S=400$, $\lambda _3={10^{ - 10}}$, $l{r_0}={0.5^{2.2}} \times {10^{ - 4}}$, $l{r_\mathrm{max}}={0.5^{2.2}} \times {10^{ - 4}}$, $l{r_\mathrm{min}}={0.5^{6.6}} \times {10^{ - 4}}$}.
\begin{algorithmic}[1]
\Statex \textbf{Input:} $\mathrm{H^*}$, $I$ and $R_1$
\Statex \textbf{Output:} ${O_{SPI - DNIO}} = {R_\theta }({R_1}({O_{DGI}}))$
\State \textbf{initialize:} randomly initialize the parameters $\theta$ in the neural network $\mathcal{R}_\theta$
\State $O_{DGI} \gets DGI(\mathrm{H^*}; I)$
\State $\tilde O \gets {R_1}({O_{DGI}})$
\For{$step = 1, 2, ..., S$}
    \State $\tilde I   \gets \mathrm{{H^*}}{R_\theta }(\tilde O)$
    \State ${{\cal L}_{\theta }} \gets \mathrm{MSE}(\tilde I,I)+{\lambda _3}\operatorname{TV} (\tilde O)$
    \State $\theta\gets \text{Adam}(\nabla_{\theta} \mathcal{L}_{\theta}, l{r_0})$
    \State $l{r_0} \gets l{r_{\min }} + \frac{1}{2}(l{r_{\max }} - l{r_{\min }})(1 + \cos (\frac{{step}}{S}\pi ))$
    
\EndFor
\end{algorithmic}

\end{algorithm}






\end{document}